\documentclass[cameraready]{Interspeech}
\usepackage{pifont}

\title{Cross-lingual Retrieval-Augmented Classification for Dysarthria Severity Assessment}

\author[affiliation={1}]{Taeyoung}{Jeong}
\author[affiliation={1}]{Insung}{Lee}
\author[affiliation={1}]{Du-Seong}{Chang}
\author[affiliation={1}, correspondingauthor]{Myoung-Wan}{Koo}

\address{
    $^1$ Department of Artificial Intelligence, Sogang University, South Korea 
}

\email{john428@sogang.ac.kr, dlstjd6474@sogang.ac.kr, dschang@sogang.ac.kr, mwkoo@sogang.ac.kr}


\usepackage{comment}

\begin{document}

\maketitle

\begin{abstract}
Automatic dysarthria severity assessment is limited by the scarcity of labeled pathological speech data. To address this, we propose Cross-lingual Retrieval-Augmented Classification (CRAC), which leverages speech from a different language via an align-retrieve-fuse pipeline. Supervised contrastive learning first shapes a severity-focused embedding space, then a vector database is built from the opposite-language corpus. During both training and inference, the classifier retrieves top-$k$ references from the aligned space and fuses them with the input via cross-attention. Evaluated on Korean post-stroke and Italian ALS dysarthria datasets under a speaker-independent three-class protocol, CRAC achieves balanced accuracies of 87.3\% on Korean and 86.7\% on Italian, improving over monolingual baselines by 8.4 and 20.0 percentage points, respectively.
\end{abstract}

\section{Introduction}
Dysarthria is a motor speech disorder caused by neurological injury or disease that degrades the motor control required for phonation, articulation, and respiration~\cite{duffy2012motor, freed2023motor}. Severity assessment is clinically essential for treatment planning and longitudinal monitoring, but traditional auditory–perceptual evaluation by speech-language pathologists (SLPs) is time-consuming, costly, and subject to inter-rater variability, motivating automatic severity assessment from speech signals~\cite{bunton2007listener}.

Early approaches combined hand-crafted features such as MFCCs, LFCCs, and prosodic descriptors with classifiers like SVMs or shallow neural networks~\cite{hasegawajohnson2006hmm, joshy2022automated}, but their performance was sensitive to feature engineering and preprocessing. More recently, self-supervised learning based pretrained encoders—including Wav2Vec~2.0~\cite{baevski2020wav2vec}, HuBERT~\cite{hsu2021hubert}, and WavLM~\cite{chen2022wavlm}—have been adopted as feature extractors, yielding substantial improvements~\cite{javanmardi2023wav2vec, sanjay2024severity, zhang2024dysarthria}. Among them, Whisper~\cite{radford2023robust} has attracted particular attention for dysarthria classification~\cite{rathod2023whisper}, as its large-scale multilingual and multi-domain pre-training provides representations that are relatively robust across diverse acoustic conditions and languages.

Despite these advances, building reliable automatic severity assessment systems remains challenging, primarily due to the scarcity of pathological speech data with clinical labels. Public corpora such as UA-Speech~\cite{kim2008dysarthric}, TORGO~\cite{rudzicz2012torgo}, and SAP~\cite{hasegawa2024community} contain limited numbers of speakers, and constructing sufficiently large clinical datasets for each language is often impractical. Cross-lingual adaptation—leveraging small labeled clinical corpora from different languages to improve performance in the target language—offers a potential solution~\cite{yeo2022cross}. However, naively pooling multilingual data can degrade performance, as models may overfit to language-specific acoustic cues rather than severity-relevant features. Contrastive learning has been explored to mitigate such domain and language mismatch~\cite{stumpf2025multilingual}, and has also shown effectiveness in improving robustness of dysarthric speech recognition~\cite{wu2021sequential}, yet practical constraints including dataset imbalance and limited clinical supervision persist.

To address these limitations, we propose a \textbf{Cross-lingual Retrieval-Augmented Classification (CRAC)} framework. Our approach is motivated by clinical practice: when assessing dysarthria severity, SLPs typically compare a patient's speech characteristics against those observed in previously diagnosed cases, drawing on accumulated clinical experience to make comparative judgments. CRAC operationalizes this reasoning process through an \emph{align–retrieve–fuse} pipeline. First, supervised contrastive alignment shapes a severity-focused embedding space that reduces cross-lingual differences. Then, at inference time, CRAC retrieves acoustically similar reference samples from a pathological speech database constructed in a different language~\cite{lewis2020retrieval}. Finally, the retrieved references are fused with the input representation to stabilize the decision boundary and mitigate domain mismatch.

We evaluate CRAC on a Korean post-stroke dysarthria dataset and an Italian Amyotrophic Lateral Sclerosis (ALS) dysarthria dataset~\cite{dubbioso2024voice} under a speaker-independent protocol, predicting subject-level severity from Maximum Phonation Time (MPT) and Diadochokinetic (DDK) tasks. Experimental results demonstrate that CRAC consistently outperforms both monolingual and naively pooled multilingual baselines—achieving balanced accuracy improvements from 66.7\% to 86.7\% on the Italian dataset and from 78.9\% to 87.3\% on the Korean dataset—while simple data pooling even degrades performance in some configurations. These findings confirm that CRAC, grounded in clinical reasoning, offers an effective strategy for dysarthric speech severity assessment under low-resource conditions.

The main contributions of this paper are as follows:
\begin{itemize}[leftmargin=*, nosep]
\item \textbf{Clinically motivated CRAC framework:} We propose CRAC, a cross-lingual retrieval-augmented classification framework inspired by the comparative reasoning process used by SLPs for severity assessment.
\item \textbf{Component-wise analysis:} We systematically analyze the effects of contrastive alignment and retrieval-based fusion via ablation studies.
\item \textbf{Cross-lingual and cross-etiology evaluation:} We evaluate CRAC on Korean post-stroke and Italian ALS dysarthria with speaker-independent testing, including top-$k$ sensitivity analysis.
\end{itemize}

\begin{figure*}[t]
\centering
\includegraphics[width=\linewidth]{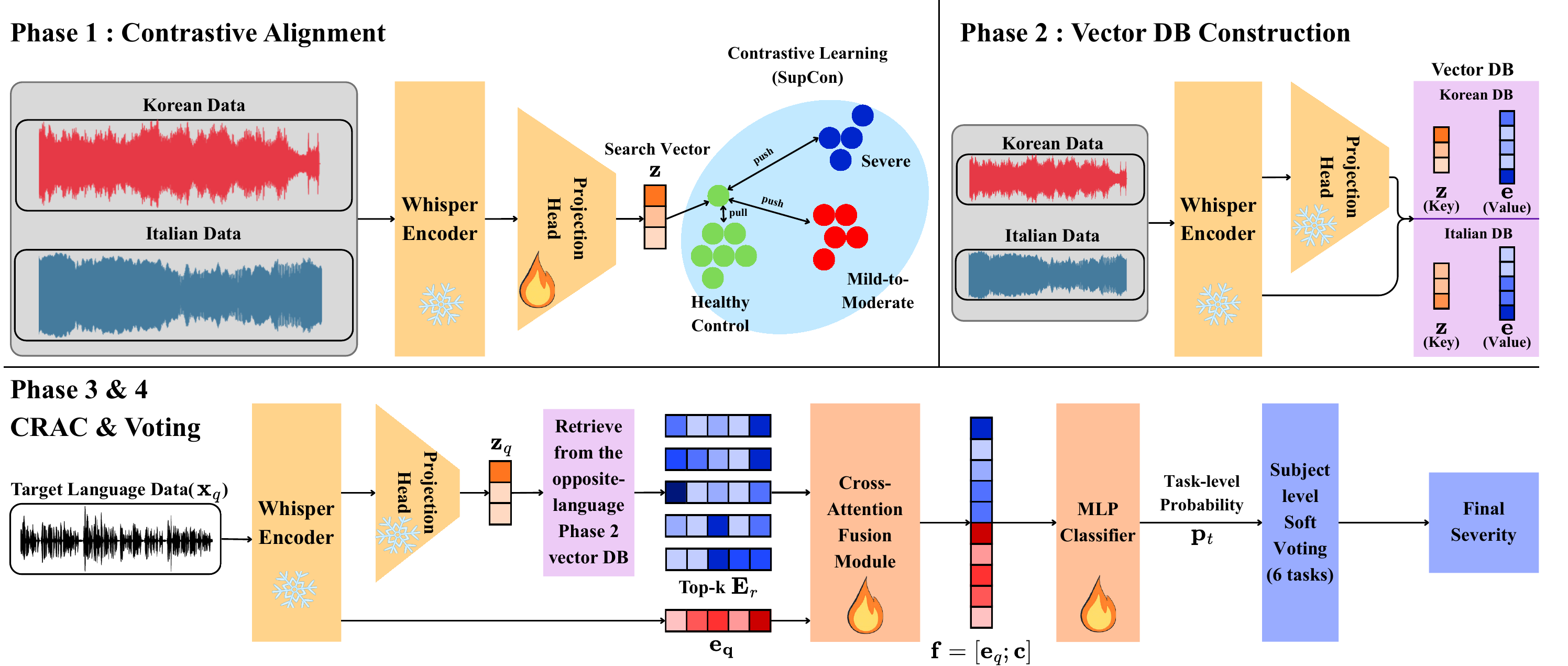}
\caption{Overview of CRAC. A projection head learns a severity-aligned search embedding $\mathbf{z}$ from Whisper features $\mathbf{e}$, and per-language vector DBs are built with keys $\mathbf{z}$ and values $\mathbf{e}$. Given a target-language query, CRAC retrieves top-$k$ references from the opposite-language DB and fuses them with $\mathbf{e}_q$ via cross-attention to form $\mathbf{f}=[\mathbf{e}_q;\mathbf{c}]$, followed by an MLP and subject-level soft voting.}
\label{fig:framework}
\end{figure*}

\section{Methods}
\vspace{-1mm}
The proposed CRAC framework operates in four phases: (1) contrastive alignment to shape a severity-focused embedding space, (2) vector database construction from a cross-lingual pathological speech corpus, (3) retrieval-augmented classification that fuses input features with retrieved references, and (4) subject-level inference via soft voting. Figure~\ref{fig:framework} illustrates the overall pipeline.

\vspace{-1mm}
\subsection{Phase 1: Contrastive Alignment}
\vspace{-1mm}
Given an input audio signal $x$, a frozen Whisper encoder $\mathcal{E}$ extracts frame-level hidden states, which are aggregated via mean pooling into a content feature vector $\mathbf{e} \in \mathbb{R}^{d}$, where $d{=}768$. A trainable projection head $g(\cdot)$, consisting of two linear layers with a ReLU activation, maps $\mathbf{e}$ to a compact search vector $\mathbf{z} = g(\mathbf{e}) \in \mathbb{R}^{m}$, where $m{=}128$, followed by L2 normalization.

To learn language-invariant, severity-focused representations, we train the projection head using supervised contrastive (SupCon) loss~\cite{khosla2020supervised}. Mini-batches are constructed by mixing samples from both languages (Korean and Italian) across all tasks (MPT and DDK) without distinction. To mitigate class imbalance in SupCon, we employ class-balanced sampling. For each anchor $i$ in a batch $I$, let $P(i)$ denote the set of indices sharing the same severity label (excluding $i$ itself), and $A(i)$ the set of all other indices. The SupCon loss is defined as:

\vspace{-4mm}
\begin{equation}
\mathcal{L}_{\text{sup}} = \sum_{i \in I} \frac{-1}{|P(i)|} \sum_{p \in P(i)} \log \frac{\exp(\text{sim}(\mathbf{z}_i, \mathbf{z}_p) / \tau)}{\sum_{a \in A(i)} \exp(\text{sim}(\mathbf{z}_i, \mathbf{z}_a) / \tau)}
\end{equation}

\noindent where $\text{sim}(\cdot, \cdot)$ denotes cosine similarity and $\tau$ is a temperature parameter. By treating all same-severity samples as positives regardless of language or task, the projection head learns to cluster samples by severity while suppressing language-specific and task-specific variations. The Whisper encoder remains frozen throughout this phase to preserve pre-trained acoustic representations. Phase~1 uses mixed-language labeled training data to learn a severity-focused search space, and Phase~2 and 3 evaluate cross-lingual retrieval augmentation by constructing the vector database from the opposite language.

\vspace{-1mm}
\subsection{Phase 2: Vector Database Construction}
\vspace{-1mm}
Using the frozen encoder and trained projection head from Phase~1, we construct a vector database from the cross-lingual pathological speech corpus (e.g., Italian data when the target is Korean, and vice versa). Each sample in the source-language corpus is encoded into a key--value pair: the L2-normalized search vector $\mathbf{z} \in \mathbb{R}^{m}$ serves as the key for efficient similarity search, while the content feature $\mathbf{e} \in \mathbb{R}^{d}$ is stored as the value to preserve rich acoustic information. The corresponding three-class severity label is also stored. This separation allows fast retrieval via cosine similarity in the compact search space while retaining high-dimensional content features for downstream fusion. To avoid any test leakage and to specifically test cross-lingual retrieval augmentation, the vector database is constructed exclusively from the opposite-language \emph{training split}. The search index is implemented using FAISS~\cite{johnson2019billion} with cosine-similarity search on normalized vectors.

\vspace{-1mm}
\subsection{Phase 3: Retrieval-Augmented Classification}
\vspace{-1mm}
During both training and inference, a target-language input $x$ is passed through the frozen encoder and projection head to obtain $\mathbf{e}_q \in \mathbb{R}^{d}$ and $\mathbf{z}_q \in \mathbb{R}^{m}$. In Phase 3, we freeze the Whisper encoder and projection head learned in Phase 1, and train only the cross-attention fusion module and Multi-Layer Perceptron (MLP) classifier. The search vector $\mathbf{z}_q$ is used to retrieve the top-$k$ most similar entries from the vector database, yielding a set of content features $\{\mathbf{e}_{r_1}, \dots, \mathbf{e}_{r_k}\} \in \mathbb{R}^{k \times d}$.

To integrate the retrieved references with the input representation, we employ a multi-head cross-attention mechanism~\cite{vaswani2017attention}. The query content feature $\mathbf{e}_q$ serves as the query, while the retrieved content features serve as both keys and values:

\vspace{-4mm}
\begin{equation}
{\small \mathbf{c} = \text{LayerNorm}\left(\text{MultiHead}(\mathbf{e}_q, \mathbf{E}_r, \mathbf{E}_r)\right)}
\end{equation}

\noindent where $\mathbf{E}_r = [\mathbf{e}_{r_1}; \dots; \mathbf{e}_{r_k}] \in \mathbb{R}^{k \times d}$ is the matrix of retrieved content features and $\mathbf{c} \in \mathbb{R}^{d}$ is the resulting context vector. The context vector is concatenated with the input content feature to form the fused representation $\mathbf{f} = [\mathbf{e}_q; \mathbf{c}] \in \mathbb{R}^{2d}$, which is fed into an MLP classifier to produce three-class severity predictions. The classifier is trained with cross-entropy loss using inverse-frequency class weights to handle class imbalance.

\vspace{-1mm}
\subsection{Phase 4: Subject-Level Inference}
\vspace{-1mm}
Each subject performs six speech tasks (three MPT: /a/, /i/, /u/; three DDK: /pa/, /ta/, /ka/). During inference, each recording is processed independently to obtain a softmax vector, and the final prediction is the argmax of the averaged probabilities across all six tasks.

\vspace{-1mm}
\section{Experimental Setup}

\vspace{-1mm}
\subsection{Datasets}
\vspace{-1mm}
We evaluate CRAC on two dysarthria datasets with distinct languages and etiologies. The \textbf{Korean (KR)} dataset consists of post-stroke dysarthric speech collected from a clinical rehabilitation setting. The \textbf{Italian (IT)} dataset~\cite{dubbioso2024voice} consists of speech from individuals with ALS. Both datasets include six speech tasks: three MPT tasks involving sustained vowel production (/a/, /i/, /u/) and three DDK tasks involving rapid syllable repetition (/pa/, /ta/, /ka/). MPT tasks assess the ability to maintain steady phonation, while DDK tasks assess articulatory speed and regularity. These tasks are commonly used in clinical dysarthria severity assessment.

Severity labels are categorized into three classes: Healthy Control, Mild-to-Moderate, and Severe. All data splits follow a speaker-independent protocol, ensuring that no speaker appears in more than one of the train, validation, and test sets. As shown in Table~\ref{tab:dataset}, both datasets exhibit notable class imbalance, with the severe class being substantially underrepresented.

\begin{table}[h]
\centering
\caption{Dataset statistics. Splits are speaker-independent. HC, Mild-to-Mod, Sev denote Healthy Control, Mild-to-Moderate, and Severe, respectively.}
\label{tab:dataset}
\footnotesize
\begin{tabular}{llcccc}
\toprule
\textbf{Data} & \textbf{Split} & \textbf{\#Files} & \textbf{\#Subj.} & \textbf{HC / Mild-to-Mod / Sev} \\
\midrule
KR & Train & 1656 & 276 & 882 / 600 / 174 \\
KR & Valid & 210  & 35  & 108 / 78 / 24 \\
KR & Test  & 210  & 35  & 108 / 78 / 24 \\
\midrule
IT & Train & 1314 & 219 & 888 / 270 / 156 \\
IT & Valid & 186  & 31  & 120 / 42 / 24 \\
IT & Test  & 132  & 22  & 90 / 30 / 12 \\
\bottomrule
\end{tabular}
\end{table}

\vspace{-1mm}
\subsection{Baselines}
\vspace{-1mm}
We compare CRAC against two baselines to isolate the effect of cross-lingual retrieval. \textbf{Baseline~1 (Monolingual)} freezes the Whisper-small encoder and trains an MLP classifier using only the target-language data, representing the standard monolingual setup. \textbf{Baseline~2 (Pooled)} trains the same architecture on pooled Korean and Italian data and evaluates on the target language; this corresponds to naive multilingual training via simple data pooling, without any explicit cross-lingual alignment or retrieval. Both baselines share the same backbone and MLP architecture; they differ only in training data composition (target-only vs pooled). 

\vspace{-1mm}
\subsection{Implementation Details}
\vspace{-1mm}
All experiments use Whisper-small~\cite{radford2023robust} with a fixed random seed of 42. In \textbf{Phase~1}, only the projection head ($768 \rightarrow 384 \rightarrow 128$) is trained using AdamW optimizer~\cite{loshchilov2017decoupled} (learning rate $5{\times}10^{-5}$, weight decay $1{\times}10^{-4}$) with cosine annealing, batch size 256, and $\tau{=}0.15$. 

In \textbf{Phase~2}, the vector database is constructed from the entire training set of the source language (Italian when the target is Korean, and vice versa) using the frozen encoder and projection head from Phase~1.


In \textbf{Phase~3}, the classifier uses top-$k{=}5$, 8 attention heads, dropout of 0.3, and MLP hidden dimensions of [512, 256]. It is trained with cross-entropy loss (inverse-frequency weights), AdamW (learning rate $1{\times}10^{-4}$, weight decay $1{\times}10^{-4}$), batch size 32.

All hyperparameters were selected using the speaker-independent validation split; the test split was used only for final evaluation.

\vspace{-1mm}
\subsection{Evaluation Protocol}
\vspace{-1mm}
Subject-level predictions are obtained via soft voting across six tasks as described in Section~2.4. We report three metrics: balanced accuracy, macro-F1, and micro-F1. Balanced accuracy, which averages per-class recall, is used as the metric to account for class imbalance. Macro-F1 provides a complementary view by equally weighting precision and recall across classes, while micro-F1 reflects overall accuracy.

\begin{table}[t]
\centering
\caption{Main comparison of subject-level severity classification. Ret.\ DB indicates the language used for the retrieval database.}
\label{tab:main}
\scriptsize
\begin{tabular}{llccc}
\toprule
\textbf{Method} & \textbf{Ret.\ DB} & \textbf{Bal-ACC} & \textbf{Macro-F1} & \textbf{Micro-F1} \\
\midrule
\multicolumn{5}{l}{\textit{Target: Korean (KR)}} \\
\midrule
Baseline 1 & -- & 0.789 & 0.800 & 0.886 \\
Baseline 2 & -- & 0.764 & 0.762 & 0.857 \\
\textbf{CRAC} & IT & \textbf{0.873} & \textbf{0.870} & \textbf{0.914} \\
\midrule
\multicolumn{5}{l}{\textit{Target: Italian (IT)}} \\
\midrule
Baseline 1 & -- & 0.667 & 0.619 & 0.773 \\
Baseline 2 & -- & 0.800 & 0.770 & 0.864 \\
\textbf{CRAC} & KR & \textbf{0.867} & \textbf{0.896} & \textbf{0.909} \\
\bottomrule
\end{tabular}
\end{table}

\section{Results}

\subsection{Main Comparison}

Table~\ref{tab:main} compares CRAC with the two baselines across both language settings. CRAC consistently achieves the best performance in all metrics. In the Korean setting, CRAC obtains a balanced accuracy of 87.3\%, improving over Baseline~1 by 8.4 percentage points (pp) and over Baseline~2 by 10.9~pp. In the Italian setting, the improvement is even more pronounced: CRAC achieves 86.7\% balanced accuracy, a 20.0~pp gain over Baseline~1.

A notable finding is that Baseline~2 does not consistently outperform Baseline~1. In the Korean setting, pooling data from both languages actually degrades balanced accuracy from 78.9\% to 76.4\%, suggesting that naive multilingual training introduces language-specific acoustic variations that act as noise rather than useful signal. While Baseline~2 does improve over Baseline~1 in the Italian setting (80.0\% vs.\ 66.7\%), CRAC still outperforms it by a substantial margin. These results indicate that simply increasing the volume of training data through cross-lingual pooling is not sufficient; rather, the structured approach of CRAC—retrieving relevant references from an aligned embedding space—provides a more effective way to leverage cross-lingual information.

\begin{table}[t]
\centering
\caption{Ablation on contrastive alignment (Align) and retrieval fusion (Ret).
Align learns the search embedding $\mathbf{z}$ (128-d).
(A) MLP($\mathbf{e}_q$); (B) MLP($\mathbf{z}_q$);
(C) Ret: FAISS key=$\mathbf{e}_q$ (no projection head) + cross-attn fusion over $\mathbf{E}_r$;
(D) full CRAC: FAISS key=$\mathbf{z}_q$ + cross-attn fusion over $\mathbf{E}_r$.}
\label{tab:ablation}
\scriptsize
\begin{tabular}{cccccc}
\toprule
& \textbf{Align} & \textbf{Ret} & \textbf{Bal-ACC} & \textbf{Macro-F1} & \textbf{Micro-F1} \\
\midrule
\multicolumn{6}{l}{\textit{Target: Korean (KR)}} \\
\midrule
A & \ding{55} & \ding{55} & 0.789 & 0.800 & 0.886 \\
B & \ding{51} & \ding{55} & 0.706 & 0.724 & 0.857 \\
C & \ding{55} & \ding{51} & 0.597 & 0.559 & 0.800 \\
D & \ding{51} & \ding{51} & \textbf{0.873} & \textbf{0.870} & \textbf{0.914} \\
\midrule
\multicolumn{6}{l}{\textit{Target: Italian (IT)}} \\
\midrule
A & \ding{55} & \ding{55} & 0.667 & 0.619 & 0.773 \\
B & \ding{51} & \ding{55} & 0.733 & 0.681 & 0.818 \\
C & \ding{55} & \ding{51} & 0.667 & 0.619 & 0.773 \\
D & \ding{51} & \ding{51} & \textbf{0.867} & \textbf{0.896} & \textbf{0.909} \\
\bottomrule
\end{tabular}
\end{table}

\subsection{Ablation Study}
Table~\ref{tab:ablation} presents the ablation results examining the individual and joint contributions of contrastive alignment and retrieval-based fusion.

Adding contrastive alignment alone~(B) yields inconsistent effects: a modest improvement for Italian (66.7\%$\rightarrow$73.3\%) but a decrease for Korean (78.9\%$\rightarrow$70.6\%). This suggests that while alignment reshapes the embedding space toward severity-focused clustering, it does not directly benefit a classifier that does not leverage cross-lingual references.

Retrieval without alignment~(C) fails to improve performance in both settings. In the Korean setting, performance drops substantially (78.9\%$\rightarrow$59.7\%), and in the Italian setting, it remains unchanged from the baseline. Without contrastive alignment, the Whisper embedding space is not organized by severity; consequently, the retrieved samples from the other language are not acoustically similar in terms of severity characteristics, providing irrelevant or misleading context to the classifier.

Only with full CRAC~(D) does performance improve substantially across both settings. This confirms that alignment and retrieval play complementary roles: contrastive alignment ensures that the retrieval operates in a severity-focused space, and retrieval provides informative cross-lingual references that enrich the classifier's input. Neither component alone is sufficient.

\subsection{Effect of Top-k}
Table~\ref{tab:topk} examines the sensitivity of CRAC to the number of retrieved references. Across both datasets, $k{=}5$ yields the best overall performance across the evaluation metrics. With $k{=}1$, the classifier has access to only a single reference, which is likely insufficient to capture the diversity of severity-related patterns. We observe a small, KR-specific dip at $k{=}3$, suggesting that performance can be sensitive to the exact composition of a small retrieved neighborhood. Performance peaks at $k{=}5$ and declines at $k{=}10$, indicating that excessive retrieval introduces noisier and less relevant samples that dilute the severity signal. Overall, these results suggest that a moderate retrieval size (around $k{=}5$) provides a favorable trade-off between informativeness and noise.

\begin{table}[t]
\centering
\caption{Effect of top-$k$ on subject-level classification.}
\label{tab:topk}
\scriptsize
\begin{tabular}{cccc}
\toprule
\textbf{Top-k} & \textbf{Bal-ACC} & \textbf{Macro-F1} & \textbf{Micro-F1} \\
\midrule
\multicolumn{4}{l}{\textit{Target: Korean (KR)}} \\
\midrule
0  & 0.789 & 0.800 & 0.886 \\
1  & \textbf{0.880} & 0.818 & 0.857 \\
3  & 0.738 & 0.753 & 0.827 \\
\textbf{5}  & 0.873 & \textbf{0.870} & \textbf{0.914} \\
10 & 0.821 & 0.804 & 0.857 \\
\midrule
\multicolumn{4}{l}{\textit{Target: Italian (IT)}} \\
\midrule
0 & 0.667 & 0.619 & 0.773 \\
1  & 0.800 & 0.770 & 0.864 \\
3  & 0.800 & 0.770 & 0.864 \\
\textbf{5}  & \textbf{0.867} & \textbf{0.896} & \textbf{0.909} \\
10 & 0.733 & 0.681 & 0.818 \\
\bottomrule
\end{tabular}
\end{table}

\begin{figure}[h]
\centering
\includegraphics[width=\linewidth]{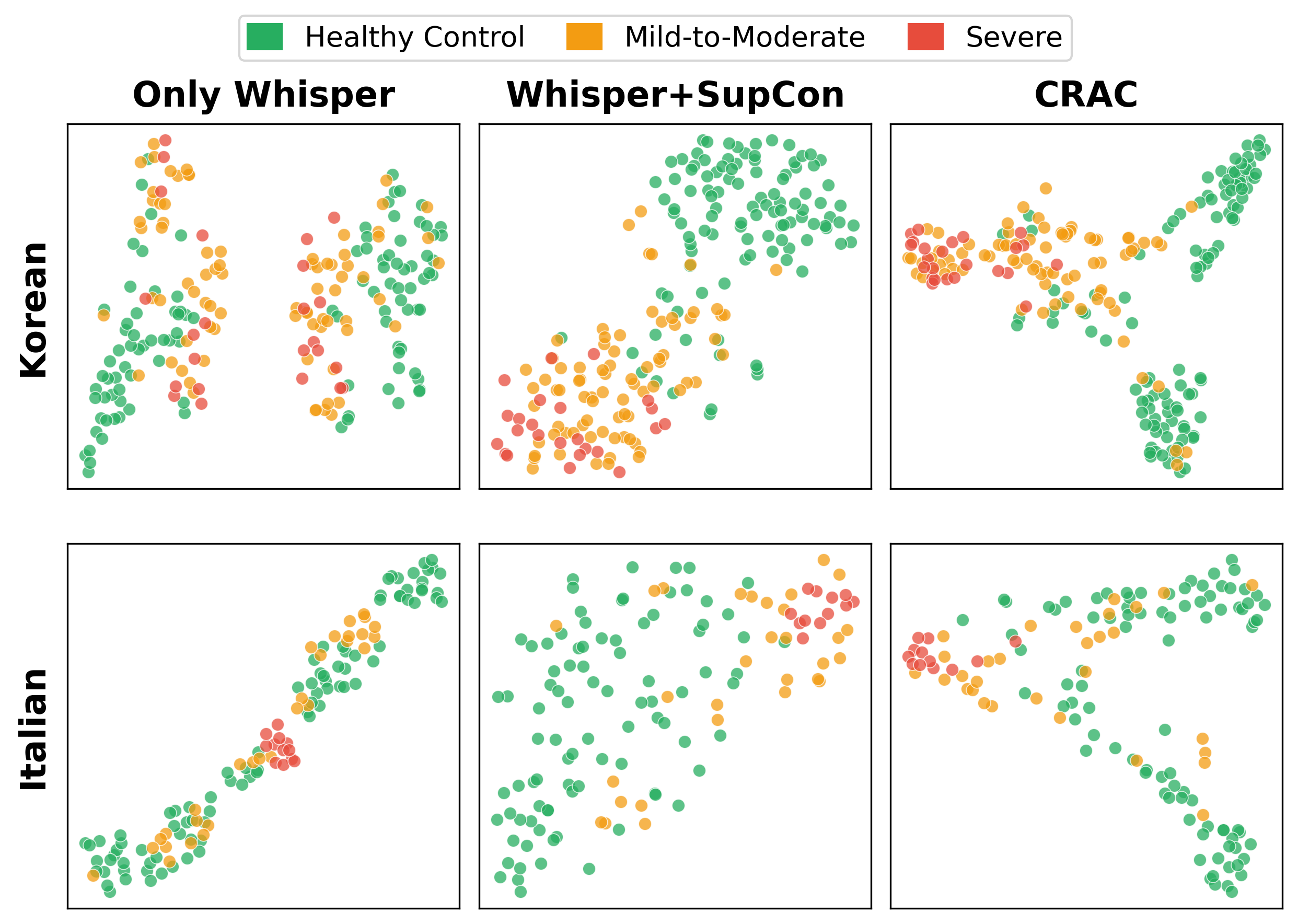}
\caption{t-SNE visualizations of test-set embeddings at three stages: content features $\mathbf{e}$ (Only Whisper), search features $\mathbf{z}$ (Whisper+Supcon), and fused features $\mathbf{f}$ (after CRAC).}
\label{fig:tsne}
\end{figure}

\subsection{Embedding Visualization}
Figure~\ref{fig:tsne} shows t-SNE projections of test embeddings at three stages of CRAC for both language settings.  Before contrastive alignment, Whisper content features ($\mathbf{e}$) exhibit substantial overlap across severity classes. After alignment, the projected search embeddings ($\mathbf{z}$) show emerging severity-wise structure, although separation between adjacent classes (e.g., healthy control vs.\ mild-to-moderate) remains ambiguous. After retrieval-augmented fusion, the fused representation ($\mathbf{f}$) yields the clearest class separation, suggesting that retrieved cross-lingual references provide complementary cues that sharpen the decision boundary. Overall, these qualitative trends are consistent with the ablation results.

\section{Conclusion}
We presented CRAC, a cross-lingual retrieval-augmented classification framework for automatic dysarthria severity assessment. Through an align--retrieve--fuse pipeline, CRAC leverages pathological speech from a different language via structured retrieval rather than naive pooling. Experiments on Korean post-stroke and Italian ALS datasets showed balanced accuracy gains of 8.4 pp and 20.0 pp over the monolingual baseline (Baseline~1), and consistently outperformed naive multilingual training via data pooling (Baseline~2). Ablation results indicate that alignment-only or retrieval-only variants do not match full CRAC performance, highlighting their complementary roles. Future work includes broader evaluation across additional languages and etiologies, along with analysis of retrieval quality and efficiency.

\section{Acknowledgments}

{This work was supported by Institute of Information \& communications Technology Planning \& Evaluation (IITP) grant funded by the Korea government(MSIT) (RS-2022- II220621, Development of artificial intelli- gence technology that provides dialog-based multi-modal explainability).}

\section{Generative AI Use Disclosure}
\textbf{GitHub Copilot} was used as an AI-assisted coding tool to support the development and editing of experimental scripts. In addition, \textbf{Gemini} was used for language editing to improve readability and grammar. In accordance with ISCA policy, all AI-assisted outputs were reviewed and, where necessary, modified by the authors, who take full responsibility for the content of this paper.


\bibliographystyle{IEEEtran}
\bibliography{mybib}

\end{document}